\documentstyle[pre,aps,epsf]{revtex}

\begin{document}

\draft


\title
{A two-species $d$-dimensional diffusive model and its mapping onto a growth model }

\author
{M. Mobilia\footnote{email:mauro.mobilia@epfl.ch} and P.-A. Bares\footnote{email:pierre-antoine.bares@epfl.ch} }

\address
{Institute of Theoretical Physics, Swiss Federal Institute of Technology of  Lausanne, CH-1015 Lausanne  EPFL, Switzerland }

\date{\today}

\maketitle

\begin{abstract}
 In this work, we consider a diffusive two-species $d-$dimensional model and study it in great details. Two types of particles, with hard-core, diffuse symmetrically and cross each other. For arbitrary dimensions, we obtain the exact density, the instantaneous, as well as non-instantaneous, two-point correlation functions for various initial conditions. We study the impact of correlations in the initial state on the dynamics. Finally, we map the one-dimensional version of the model under consideration onto a growth model of RSOS type with three states and solve its dynamics.
   
\end{abstract}
\pacs{PACS number(s): 02.50.Ey, 68.35.Fx, 66.30.-h, 05.50.+q}
\section{Introduction}
Stochastic reaction-diffusion models have attracted recently much interest in  (see e.g. \cite{Privman,Schutz0} and references therein). In the last decade, the latter appeared (directly or via mapping) as models for traffic flow \cite{Schadschneider}, kinetic biopolymerization\cite{Macdonald}, reptation of DNA in gels \cite{Widom}, interface growth \cite{Krug,Episov}, etc.

In this context, simple symmetric (SEP) \cite{Stinchcombe0} and asymmetric exclusion processes, in one dimension, 
(ASEP) \cite{Privman,Schutz0,Krug1} play a particular role because of their relationships with integrable quantum spin systems (Heisenberg chains) and because of their connection with the KPZ equation \cite{Krug2}, directed polymers in random media \cite{Krug}, and shock formation (see e.g. \cite{Privman} and references therein). These models have been extensively studied and the ASEP with open boundary conditions, as a simple driven diffusion model, exhibits a rich dynamical behaviour involving different nonequilibrium phase transitions in the steady states. They can be studied exactly on the basis of the so called {\it Matrix Approach} (MA), an algebraic approach based on an ansatz for the probability distribution which is related to the integrability of some quantum spin chains. This approach provides the full solution of the ASEP (and also the SEP) model, including the full phase diagram, density profile and, in principle, any equal time correlation functions. Though, only few explicit results are  known \cite{Schutz1} about the dynamical correlation functions, much work has been done on the static properties. 

The MA  has been generalized to solve the stationary states of one-dimensional  models with several species \cite{Derrida} and, recently, a first-order phase-transition in  some models has been found \cite{Arndt} (see also \cite{Rajewsky} where different results were obtained, independently, for the same model).

The lack of exact results for the dynamics of multispecies models \cite{Privman1,Belitsky} (in particular in dimension $d>1$ see also \cite{BM} and references therein), has motivated us to study in some details the {\it dynamics} of a two-species model, which is related to the models introduced by Arndt {\it et al.} in \cite{Derrida,Arndt}. We compute explicitly, in arbitrary dimensions, the density, and the two-point instantaneous/non-instantaneous  correlation functions.
We then exploit the exact results for the two-point correlation function to study a RSOS-type \cite{Krug,Episov,Grynberg,Grynberg1} growth model.

The paper is organized as follows: In section II, we present the general {\it stochastic} formalism within which we will work.
 In section III, for the model under consideration, we compute in arbitrary dimensions  the density for various initial states and in presence/absence of initial correlations.
 In section IV, for a translationally invariant version of our model, we evaluate the instantaneous two-point correlation functions in arbitrary dimensions. In particular, in one-spatial dimension, we assume both cases where initial correlations are absent/present. 
In section V,  we introduce and solve a growth model of RSOS-type ``with three states''. This analysis is carried out for different initial states (correlated and uncorrelated).
Finally, in section VI, we calculate for systems with  uncorrelated (but random) as well as correlated initial states the non-instantaneous two-point correlation functions.

\section{The Formalism and the Model}
Consider an hypercubic lattice of dimension $d$ with $N$ sites ($N=L^d$), where 
$L$ represents the linear dimension of the hypercube, and periodic boundary conditions are imposed.
Further assume that local bimolecular reactions between species $A$ and $B$ 
take place. Each site is either empty (denoted by the symbol $0$)  
or occupied at most by one particle of type $A$ (respectively $B$) 
denoted in the following by the
index $1$ (respectively $2$). 
The dynamics is parametrized by the transition rates
$\Gamma_{\alpha \beta}^{\gamma \delta}$, where $\alpha,\beta,\gamma,
\delta=0,1,2$: $\forall (\alpha, \beta)\neq (\gamma, \delta)\;\;,\;\;  
\Gamma_{\alpha \beta}^{\gamma \delta}\;: \;\alpha + \beta \longrightarrow 
\gamma +\delta$.

Probability conservation implies $\Gamma_{\alpha \beta}^{\alpha \beta}=-\sum_{(\alpha,\beta)\neq(\alpha',\beta')}
\Gamma_{\alpha \beta}^{\alpha' \beta'},$
with $\Gamma_{\alpha \beta}^{\gamma \delta}\geq 0, \forall (\alpha,\beta)\neq(\gamma,\delta)$.

For example the rate $\Gamma_{2 2}^{1 2}$ corresponds to the process
$B B \longrightarrow A B$, while conservation of probability leads to 
$\Gamma_{1 1}^{1 1}=-(\Gamma^{1 0}_{1 1}+\Gamma^{0 1}_{1 1}+
\Gamma^{0 0}_{1 1}+\Gamma^{0 2}_{1 1}+
\Gamma^{2 0}_{1 1}+\Gamma^{2 1}_{1 1}+ \Gamma^{1 2}_{1 1}+ 
\Gamma^{2 2}_{1 1})$

The state of the system is determined by specifying the probability for the
occurence of
configuration $\{ n\}$ at time $t$. It is represented by the ket $|P(t)\rangle=\sum_{\{n\}}P(\{n\},t)|n \rangle$, where the sum runs over the $3^N$ configurations ($N=L^d$).
At site $i$ the local state is denoted by the ket 
$|n_i\rangle =(1 \; 0 \; 0)^{T}$ if the site 
$i$ is empty,  $|n_i\rangle =(0 \; 1 \; 0)^{T}$ if the site $i$ is occupied by 
a particle of type $A$ 
($1$) and  $|n_i\rangle =(0 \; 0 \; 1)^{T}$ 
otherwise.
It is by now well established that the Master equation governing the dynamics of the systems can be rewritten as an imaginary-time Schr\"odinger equation:
\begin{eqnarray}
\label{eq.0.3.1}
\frac{\partial}{\partial t}|P(t)\rangle=-H |P(t)\rangle,
\end{eqnarray}
where $H$ denotes the Markov generator, also called {\it stochastic Hamiltonian}, and is in  general neither hermitian nor normal. Its explicit form is given below.
We also introduce the {\it left vacuum} $\langle \widetilde \chi|$ which is defined by
\begin{eqnarray}
\label{eq.0.4}
\langle \widetilde \chi|\equiv \sum_{\{n\}} \langle\{n\} |
\end{eqnarray}
Probability conservation yields the local equation (stochasticity of $H$):
$\langle \widetilde \chi|H = \sum_{e^{\alpha}}\sum_{{\bf m}}\langle \widetilde 
\chi|H_{{\bf m}, {\bf m}+e^{\alpha}}=0 \Longrightarrow \langle \widetilde 
\chi|H_{{\bf m}, {\bf m}+e^{\alpha}}=0, $

where $e^{\alpha}$  denotes the unit vector in the direction $\alpha$ ($1\leq\alpha\leq d$) and ${\bf m}$ designates a point of the hyperlattice labelled with help of its $d$ components: ${\bf m}=(m_{1},\dots,m_{d})$ .

In this work, we assume that there are  only  symmetric  nearest-neighbour jump-processes. A particle $A$ (respectively $B$) can jump, with rate $\Gamma_{0 1}^{1 0}=\Gamma_{1 0}^{0 1}>0 $ (respectively $\Gamma_{0 2}^{2 0}=\Gamma_{2 0}^{0 2}>0$) to an adjacent site (in the $d$ directions) if the latter was previously empty. Such processes are symbolized by the ``reaction'' $A\emptyset \longleftrightarrow \emptyset A $ (respectively, $B\emptyset \longleftrightarrow \emptyset B $ ). In addition we assume that when two different particles $A$ and $B$ are adjacent, they can cross each other with rate $\Gamma_{1 2}^{2 1}= \Gamma^{1 2}_{2 1}>0$. This processes are schematized by the reaction $A B\longleftrightarrow B A $.

The local Markov generator corresponding to this system and which  acts on two adjacent sites ${\bf m}$ and ${\bf m}+e^{\alpha}$ reads  
\begin{eqnarray}
\label{eq.0.5.2}
-H_{{\bf m},{\bf m}+e^{\alpha}} =
\left(
 \begin{array}{c c c c c c c c c}
 0 & 0  & 0 & 0  & 0  & 0 & 0 & 0 & 0 \\
 0  & \Gamma_{0 1}^{0 1} & 0 &\Gamma_{1 0}^{0 1} & 0  & 0 & 0 & 0 & 0 \\
 0  & 0 &\Gamma_{0 2}^{0 2}& 0 & 0 & 0 &  \Gamma_{2 0}^{0 2} & 0 & 0 \\
 0 & \Gamma_{0 1}^{1 0} & 0 &\Gamma_{1 0}^{1 0} & 0  & 0  & 0 & 0 & 0 \\
0 & 0 & 0 & 0 & 0 & 0  & 0 & 0 & 0 \\
 0 & 0 & 0& 0 & 0 & \Gamma_{1 2}^{1 2} & 0 & \Gamma_{2 1}^{1 2} & 0 \\
0 & 0 &\Gamma_{0 2}^{2 0}& 0 & 0 & 0&  \Gamma_{2 0}^{2 0} & 0 & 0 \\
0 & 0 & 0& 0 & 0 & \Gamma_{1 2}^{2 1} & 0 & \Gamma_{2 1}^{2 1} & 0\\
 0 & 0 & 0& 0 & 0 & 0 & 0 & 0& 0 \\
 \end{array}\right)\
\end{eqnarray}
where the same notations as in reference \cite{Fujii,BM} have been used.
Probability conservation implies that each column in the above representation
sums up to zero. 

Locally, the left vacuum $\langle \widetilde \chi |$ has the representation
$\langle \widetilde \chi |=(1\,  1\,  1)\otimes (1 \, 1 \, 1)$
The action of any operator on the left-vacuum has a simple summation 
interpretation. This observation will be crucial in the following.
Below we shall assume an initial state $|P(0)\rangle$ and investigate the
expectation value of an operator $O$ (observables such as density etc.) :
$\langle O \rangle(t)\equiv \langle \widetilde \chi|O e^{-Ht}|P(0)\rangle$

From (\ref{eq.0.5.2}), we can compute the equations of motion of the density and of the two-point correlation functions \cite{BM}.
For the density, we have:
\begin{eqnarray}
\label{eq.0.7}
\frac{d}{dt}\langle n_{\bf m}^{A,B}(t)\rangle \equiv
\frac{d}{dt}\langle \widetilde \chi| n_{\bf m}^{A,B}e^{-Ht}|P(0)\rangle=
-\sum_{e^{\alpha}}\langle n_{\bf m}^{A,B}(H_{{\bf m},{\bf m}+e^{\alpha}}+ H_{{\bf m}-e^{\alpha},{\bf m}})\rangle(t)
\end{eqnarray}

For the derivation of the equation of motion of two-point correlation functions we would proceed similarly, as in \cite{BM}. One should however pay attention to distinguish the case of the correlation function of adjacent sites from the general case.

In general, when $\Gamma_{10}^{01}$, $ \Gamma_{01}^{10}$, $\Gamma_{20}^{02}$, $ \Gamma_{02}^{20}$,  $\Gamma_{12}^{21}$, $ \Gamma_{21}^{12}$ are independent parameters the equations of motions of the multi-point correlation functions
 constitute an open hierarchy and the {\it dynamics} is not soluble.
 The stationary states of such systems have been studied in \cite{Derrida} by Arndt {\it et al.} Recently it has been  shown \cite{Arndt} with help of {\it quadratic algebra} techniques \cite{Arndt,Derrida} and numerical means that an asymmetric version of this  model
exhibits a  first-order phase transition, in its stationary state, when 
$\Gamma_{10}^{01}=\Gamma_{01}^{10}=\Gamma_{20}^{02}=\Gamma_{02}^{20}
=\Gamma_{12}^{21}=1$ and $\Gamma_{12}^{21}=q$. 
The steady-state of the density of the same model has also been studied 
independently  by Rajewsky {\it et al.} \cite{Rajewsky} who obtained different results: in \cite{Rajewsky}, the authors, argued that there is no phase transition from the ``mixed phase'' to a ``disordered phase''.
\\

Here we assume
\begin{eqnarray}
\label{eq.0.10}
\Gamma_{10}^{01}= \Gamma_{01}^{10}=\Gamma_{20}^{02}=\Gamma_{02}^{20}=
\Gamma_{12}^{21}= \Gamma_{21}^{12}=\Gamma,
\end{eqnarray}
which guarantees that the equations of motion of the correlation functions close in arbitrary dimensions. 

From now on we focus on the soluble model described by (\ref{eq.0.5.2})
with equal rates, according to the {\it solubility constraints} (\ref{eq.0.10}).

Before studying statistical and dynamical properties of this model, let us comment on its solvability.
In the single species reaction-diffusion models, the solvability inherent to the closure of the hierarchy was explained in the framework of the duality transformations. In fact, it has been shown that the spectrum of the single-species stochastic Hamiltonian (with the solubility constraints) is identical to the spectrum of  an anisotropic spin-$1/2$ Heisenberg quantum Hamiltonian $H_{XXZ}$ in a magnetic field \cite{Schutz2}.
As shown in  \cite{Fujii}, the situation is quite different for the multispecies problem and  a general, comprehensive and unified understanding of the formal solubility is still lacking. However, for the model under consideration here, it has been shown \cite{Alcaraz} that the stochastic Hamiltonian (\ref{eq.0.5.2}), can be mapped, via a similarity transformation, to an exactly integrable quantum  spin-1
model introduced by Sutherland \cite{Sutherland}.
\section{Exact study of the density}
In this section, we study the density of the 
system, in particular, when translation invariance is broken (the initial density is non-uniform) and when correlations in the initial state are present. 

It follows from (\ref{eq.0.7}) that the density of species $j\in(A,B)$ at site ${\bf m}$, labelled with its $d$ components (${\bf m}=(m_{1},\dots,m_{d})$), obeys to the following linear differential-difference equation \cite{BM}:

\begin{eqnarray}
\label{eq.5.2}
\frac{d}{dt}\langle n_{{\bf m}}^{j}\rangle(t)=
-2\Gamma d \langle n_{{\bf m}}^{j}\rangle(t)+\Gamma \sum_{\alpha=1}^{d}\left(\langle n_{{\bf m}+e^{\alpha}}^{j}\rangle  + \langle n_{{\bf m}-e^{\alpha}}^{j}\rangle\right)
\end{eqnarray}

We first consider the situation where particles are initially non-uniformally distributed.
Namely, we assume that
particles of type $B$ are located in the region of space 
$L/2<x_{1}\leq L, \dots, L/2<x_{d}\leq L$ while particles of
type $A$ are initially 
confined in the region $0\leq x_{1}\leq L/2, \dots,  
0\leq x_{d}\leq L/2$ (we assume that $L$ is even). Within each of the two regions, particles of each type 
are distributed
uniformally with respective densities $\rho_{B}(0)$ and  $\rho_{A}(0)$.
Solving (\ref{eq.5.2}) for this initial condition, we find:
\begin{eqnarray}
\label{eq.5.5}
\langle n_{{\bf m}}^{A}(t)\rangle&=&\rho_{A}(0)e^{-2d\Gamma t}\sum_{{\bf m'}}\langle n_{{\bf m'}}^{A}(0)\rangle \prod_{\alpha=1}^{d}I_{m_{\alpha}-m_{\alpha}' }(2\Gamma t)=\rho_{A}(0)e^{-2d\Gamma t}\sum_{{\bf m'}} \prod_{\alpha=1}^{d}[\Theta\left(\frac{L}{2}-m_{\alpha}\right)I_{m_{\alpha}-m_{\alpha}' }(2\Gamma t)]\nonumber\\&=&\rho_{A}(0)\prod_{\alpha=1}^{d}\left(\sum_{0\leq m'_{\alpha}\leq \frac{L}{2}}e^{-2\Gamma t}I_{m_{\alpha}-m_{\alpha}'}(2\Gamma t)\right),
\end{eqnarray}
where $I_{n}(z)$  denotes the usual modified Bessel function. We have also introduced the Heaviside function  $\Theta(x)=0$ if $x<0$ and  $\Theta(x)=1$ if $x>1$.

With help of the asymptotic expansion of the Bessel functions  and approximating $\sum_{ n_{\alpha}}e^{-\frac{n_{\alpha}^{2}}{4\Gamma t}}\approx \int dn_{\alpha}e^{-\frac{n_{\alpha}^{2}}{4\Gamma t}} $, we obtain from (\ref{eq.5.5}) the long-time behavior of the density at site ${\bf m}$:
\begin{eqnarray}
\label{eq.5.5}
\langle n_{{\bf m}}^{A}(t)\rangle\approx
\rho_{A}(0)\prod_{\alpha=1}^{d}\left[\frac{erf\left(\frac{L/2-m_{\alpha}}{\sqrt{4\Gamma t}}\right)-erf\left(\frac{m_{\alpha}}{\sqrt{4\Gamma t}}\right) }{2}\right],
\end{eqnarray}
where $erf(z)$  denotes the usual error function. 

Similarly, we have for the density of particles $B$:
\begin{eqnarray}
\label{eq.5.6}
\langle n_{{\bf m}}^{B}(t)\rangle&=&\rho_{B}(0)\prod_{\alpha=1}^{d}\left(\sum_{\frac{L}{2}< m'_{\alpha}\leq L}e^{-2\Gamma t}I_{m_{\alpha}-m_{\alpha}'}(2\Gamma t)\right)\approx
\rho_{B}(0)\prod_{\alpha=1}^{d}\left[\frac{erf\left(\frac{L-m_{\alpha}}{\sqrt{4\Gamma t}}\right)-erf\left(\frac{L/2-m_{\alpha}}{\sqrt{4\Gamma t}}\right) }{2}\right]
\end{eqnarray}

 We now pass to the case where the
distribution of particles for each species
$j\in(A,B)$ is given by ($\kappa_{j}$ denotes a real dimensionless constant) an initially correlated distribution:
\begin{eqnarray}
\label{eq.5.9.0}
\langle n_{{\bf m}}^{j}\rangle (0)=\rho_{j}(0)\left(\prod_{i=1\dots d}\delta_{m_{i},0}+\kappa_{j}\prod_{i=1\dots d}|m_{i}|^{-\gamma_{i}}\left(1-\delta_{m_{i},0}\right)\right),
\end{eqnarray}
The exact densities then read \cite{BM}
\begin{eqnarray}
\label{eq.5.9}
\langle n_{{\bf m}}^{j}\rangle (t)=\rho_{j}(0)\left[\prod_{i=1 \dots d}
\left(e^{-4\Gamma t}I_{m_{i}}(2\Gamma t)\right)+\kappa_{j} \prod_{i=1 
\dots d}\left( e^{-2\Gamma t}\sum_{m_{i}'\neq 0} |m_{i}'|^{-\gamma_{i}} 
I_{m_{i}-m_{i}'}(2\Gamma t)\right)\right]
\end{eqnarray}
For $\kappa_{j}\neq 0$, in the limit $m\sim L\gg 1$ and
$\Gamma t \gg1$, with  $\sigma\equiv m/L $ and  $u\equiv \frac{L^2}{4\Gamma t}$. When $\sigma={\cal O}(1)$, then $u\sim\frac{\sigma^2 L^2}{4\Gamma t}=\frac{m^2}{4\Gamma t}$,  we obtain
\begin{eqnarray}
\label{eq.5.10}
\langle n_{{\bf m}}^{j}\rangle (t)\sim
\left\{
  \begin{array}{l l l}
 \prod_{i=1}^{d}\frac{2e^{-\sigma^{2}u \sqrt{u\sigma^{2}/\pi}}}
{(1-\gamma_{i})(4u\sigma^{2}\Gamma t)^{\gamma_{i}/2}} &\mbox{if $0\leq\gamma_{i}<1$} \\
 \prod_{i=1}^{d} \frac{(1+e^{-\sigma^{2}u})\zeta(\gamma_{i})}{(\pi \Gamma t)^{1/2}}   &\mbox{if $\gamma_{i} >1$}\\
 \frac{(e^{-\sigma^{2}u}\ln(4u\sigma\Gamma t))^{d}}{(\pi\Gamma t)^{d/2}} &\mbox{if$\gamma_{i} =1$},
\end{array}
\right.
\end{eqnarray}
where $\zeta(\nu)=\sum_{k\geq 1}k^{-\nu}\;, \nu>1$ is the Riemann zeta function.

It follows from these results that for $\gamma_{i}\neq 1$, the density decays as a power-law of time. However, notice that when  initial correlations are {\it strong} (i.e. $0<\gamma_{i}<1$), the algebraic decay of the density is {\it non-universal} (it depends of $\gamma_{i}$). When  initial correlations are {\it weak} (i.e. $\gamma_{i}>1$), the algebraic decay of the density is {\it universal}. Hence the case where $\gamma_{i}=1$ is {\it marginal} and there are logarithmic corrections to the {\it universal behaviour}.

For initial states decribed by (\ref{eq.5.9.0}), with $\gamma_{i}=0$ and $\kappa_{j}\neq 0$, then we have
\begin{eqnarray}
\label{eq.5.10.1}
\langle n_{{\bf m}}^{j}\rangle (t)\sim
\rho_{j}(0) \left(\frac{e^{-d\sigma^{2}u}}{(4\pi\Gamma t)^{d/2}}+\frac{\Gamma(1/2)-\Gamma(1/2,u)}{\sqrt{4\pi}}(1+\frac{1}{8\Gamma t})\right)
\end{eqnarray}
Therefore the dimensionality has a non-trivial effect: when $d<2$, the densities decays as $t^{-d/2}$. Otherwise, when  $d>2$, $\langle n_{\bf m}^{j}\rangle (t)\sim t^{-1}$.

On the other hand, when $\kappa_{j}=0$,
the initial density of species $j$ vanishes on the hypercube
except at the origin, where its value is $\rho_{j}(0)$.
In this case, the limit considered above, yields
\begin{eqnarray}
\label{eq.5.11}
\langle n_{{\bf m}}^{j}\rangle (0)\sim
\frac{e^{-d\sigma^2 u}}{(4\pi\Gamma t)^{d/2}}
\end{eqnarray}
Notice that because of conservation of the number of particles, in the {\it translationally invariant situation}, we  simply have
\begin{eqnarray}
\label{eq.5.4}
\rho_{A}(t)=\rho_{A}(t=0)\equiv\rho_{A}\;\;;
\rho_{B}(t)=\rho_{B}(t=0)\equiv\rho_{B}
\end{eqnarray}
\section{Instantaneous Two-point Correlation functions for 
translationally invariant systems}
In this section we compute exactly the two-point  correlation function for translationally invariant systems, in arbitrary dimensions for different initial states.

The equations of motion for the connected correlation functions
${\cal C}_{{\bf r}}^{ij}(t)\equiv{\cal C}_{-{\bf r}}^{ij}(t) \equiv
\langle n_{{\bf l}}^{i}n_{{\bf m}}^{j}\rangle(t)-\rho_{i}
\rho_{j}
=\langle n_{0}^{i}n_{{\bf m-l}}^{j}\rangle(t)-\rho_{i}
\rho_{j}, (i,j)\in (A,B)$, read, with the notation: ${\bf r}=(r_{1},\dots,r_{\alpha},\dots r_{d})\equiv{\bf m-l}$, where $\alpha=1,\dots,d$.
\begin{eqnarray}
\label{eq.5.8}
&&\frac{\partial}{\partial t}{\cal C}_{{\bf r}}^{ij}(t)=-4\Gamma d 
 {\cal C}_{{\bf r}}^{ij}(t)
+2\Gamma \sum_{\alpha=1}^{d}\left({\cal C}_{{\bf r}+e^{\alpha}}^{ij}(t) +{\cal C}_{{\bf r}-e^{\alpha}}^{ij} (t) \right), \; ||{\bf r}||\geq 2, \nonumber\\
&&\frac{\partial}{\partial t}{\cal C}_{e^{\alpha}}^{ij}(t)=2\Gamma\left[{\cal C}_{2e^{\alpha}}^{ij}(t)
+\sum_{\alpha'\neq \alpha=1\dots d}\left\{{\cal C}_{e^{\alpha}-e^{\alpha'}}^{ij}(t)+ {\cal C}_{e^{\alpha}+e^{\alpha'}}^{ij}(t) \right\} -(2d-1){\cal C}_{e^{\alpha}}^{ij}(t)  \right], \nonumber\\
&&\frac{\partial}{\partial t}{\cal C}_{0}^{ij}(t)=0
\end{eqnarray}
Solving the latter, we have, using known properties of modified Bessel functions (see appendix,(\ref{eq.A.1})):
\begin{eqnarray}
\label{eq.5.9}
{\cal C}_{{\bf r}}^{ij}(t)&=&\sum_{{\bf r'}\neq 0}{\cal C}_{{\bf r'}}^{ij}(0) 
e^{-4d\Gamma t}\prod_{\alpha=1}^{d} I_{r_{\alpha}-r'_{\alpha}}(4\Gamma t)
+{\cal C}_{0}^{ij}(0) e^{-4d\Gamma t}\prod_{\alpha=1}^{d}I_{r_{\alpha}}(4\Gamma t)
 \nonumber\\
&-&\int_{0}^{t} d\tau e^{-4d \Gamma  \tau}  {\cal C}_{e^{\alpha}}^{ij}(t-\tau) \left(\frac{\partial}{\partial \tau}-4d{\Gamma}\right)\prod_{\alpha=1}^{d} I_{r_{\alpha}}(4\Gamma \tau)-4d\Gamma\int_{0}^{t}d\tau e^{-4d\Gamma\tau}{\cal C}_{e^{\alpha}}^{ij}(t-\tau)\prod_{\alpha=1}^{d}I_{r_{\alpha}}(4\Gamma \tau)
\nonumber\\
&+&2\Gamma \int_{0}^{t}d\tau e^{-4d\Gamma \tau}\sum_{\alpha=1}^{d}{\cal C}_{e^{\alpha}}(t-\tau)\left[I_{r_{\alpha}+1}(4\Gamma \tau)+I_{r_{\alpha}-1}(4\Gamma \tau)\right]\prod_{\alpha'\neq\alpha}I_{r_{\alpha'}}(4\Gamma \tau)
\end{eqnarray}
Restricting the solution to {\it one-spatial dimension}, with $r\equiv |m-l|\geq 0$, the Laplace transform yields ($\; i,j \in (A, B)$)
\begin{eqnarray}
\label{eq.5.10}
{\cal C}_{1}^{ij}(t)=e^{-4\Gamma t}\sum_{r'\geq 1}{\cal C}_{r'}^{ij}(0)\left\{I_{r'}(4\Gamma t)  + I_{r'-1}(4\Gamma t) \right\}
\end{eqnarray}
and more generally,
\begin{eqnarray}
\label{eq.5.10.0}
{\cal C}_{r\geq 1}^{ij}(t)=e^{-4\Gamma t}\sum_{r'\geq 1}{\cal C}_{r'}^{ij}(0)\left\{I_{r'+r-1}(4\Gamma t)  + I_{r'-r}(4\Gamma t) \right\}
\end{eqnarray}
Let us now consider {\it one-spatial dimension} and assume that the initial correlations are given by
\begin{eqnarray}
\label{eq.5.11}
{\cal C}_{r}^{l}(0)=\kappa_{l} r^{-\nu_{l}}, \; \nu_{l}\geq 0, 
\; l\in (AA,BB, AB)
\end{eqnarray}
We discuss the case $|\kappa_{l}|>0$
while the case $\kappa_{l}=0$ corresponds either to
the situation where no particle is present on the lattice initially, or,
when all sites of the lattice are occupied by particles of species
$i$ (or $j$). An alternative is  that the system would be initially in its steady-state.
When a single species is present initially, say species $A$, we recover the known problem of symmetric diffusion of hard particles 
$A+\emptyset \leftrightarrow  \emptyset + A$. When the lattice is full (or empty) initially, no dynamics takes place. A single-species one-dimensional process $A+A\leftrightarrow \emptyset+\emptyset $ with a correlated initial state as in (\ref{eq.5.11}) has been studied in \cite{Grynberg}.

Again we can infer the asymptotic behavior of the two-point 
connected correlation functions
in the limit $\Gamma t \gg 1$ with $v\equiv \frac{L^2}{8\Gamma t} <\infty$ \cite{BM}.

It is  useful for the sequel to introduce  the definitions of the auxiliary functions \cite{BM}:

\begin{eqnarray}
\label{eq.5.14.1}
{\cal F}_{1}(v,\sigma, \nu_{l})\equiv \frac{ \left(\Gamma(\frac{1-\nu_{l}}{2}) +\Gamma(\frac{1-\nu_{l}}{2},\sigma^2 v)
-\Gamma(\frac{1-\nu_{l}}{2},v(1-\sigma)^2)-
 \Gamma(\frac{1-\nu_{l}}{2},v(1+\sigma)^2)\right) }{\sqrt{4\pi}} 
\end{eqnarray}
\begin{eqnarray}
\label{eq.5.14.2}
{\cal F}_{2}(v,\sigma, \nu_{l})\equiv \frac{e^{-\sigma^2 v }}{1-\nu_{l}}\sqrt{\frac{v\sigma^2}{\pi}} 
\end{eqnarray}

We distinguish two regimes\\
 
i) For $r\ll L$, with $r_{\alpha}^2/8\Gamma t \ll 1$ and $\sigma
\equiv \frac{r}{L}$, 
\begin{eqnarray}
\label{eq.5.12}
{\cal C}_{r}^{l}(t)\sim 
\left\{
  \begin{array}{l l l}
     \frac{\kappa_{l} {\cal F}_{1}(v,\sigma,\nu_{l}) }
{\sqrt{4\pi}(8\Gamma t)^{\nu_{l}/2}}
&\mbox{if $0\leq\nu_{l}<1$}\\
     \frac{\kappa_{l} \left(2\zeta (\nu_{l}) +(8v\Gamma \sigma_{\alpha}^2 t)^{\frac{1-\nu_{l}}{2}}\right) }{(8\pi\Gamma t)^{1/2}} &\mbox{if $\nu_{l} >1$}\\
 \frac{\kappa_{l}\ln{(8\Gamma v(1-\sigma_{\alpha})t)}}{(8\pi\Gamma t)^{1/2}}
  &\mbox{if $\nu_{l} =1$},
\end{array}
\right.
\end{eqnarray}
ii) For $r\gg 1, r\equiv \sigma L \sim L$, we have

\begin{eqnarray}
\label{eq.5.13}
{\cal C}_{r}^{l}(t)\sim 
\left\{
  \begin{array}{l l l}
  \kappa_{l} \left( \frac{{\cal F}_{2}(v,\sigma,\nu_{l})}{\sqrt{4\pi}
  (8\Gamma t)^{\nu_{l}/2}} \right)
&\mbox{if $0\leq\nu_{l}<1$} \\
   \frac{ \kappa_{l}\left[(1+e^{-\sigma^{2}v})\zeta(\nu_{l})+((1-\sigma)
   /\sigma)(8v\Gamma \sigma^2 t)^{\frac{1-\nu_{l}}{2}} \right]  }
   {(8\pi\Gamma t)^{1/2}} &\mbox{if $\nu_{l} >1$}\\
 \frac{\kappa_{l} e^{-\sigma^{2}v}\ln(8\Gamma v\sigma t)}{(8\pi\Gamma t)^{1/2}}
  &\mbox{if
 $\nu_{l} =1$},
\end{array}
\right.
\end{eqnarray}

As for the density, it follows from these results that for $\nu_{l}\neq 1$, the (connected-)correlation functions  decay as a power-law of time. When initial correlations are {\it strong} (i.e. $0<\nu_{l}<1$) the power-law decay of the correlation functions is {\it non-universal} . In contrast, when  initial correlations are {\it weak} (i.e. $\nu_{l}>1$), the algebraic decay of the  correlation functions  is {\it universal}. The case where $\nu_{l}=1$ is {\it marginal} and  logarithmic corrections to the {\it universal behaviour} arise.

In arbitrary dimension ($d\geq 1$), we consider a translationally invariant random but uncorrelated intial state, described by:
\begin{eqnarray}
\label{eq.5.14}
{\cal C}_{{\bf r}}^{l}(0)=\kappa_{l},  l\in (AA,BB, AB),
\end{eqnarray}
where, as above, $\kappa_{l}\neq 0$.

The asymptotic behaviour
 ($L, \Gamma t \gg 1$ with $v\equiv L^2/8\Gamma t <\infty$ and  $\sigma_{\alpha}\equiv r_{\alpha}/L$) of the connected correlation functions is then \cite{BM}:

\begin{eqnarray}
\label{eq.3.35.1}
{\cal C}_{{\bf r}}^{l}(t)\sim 
\left\{
  \begin{array}{l l}
\kappa_{l}\left(1+\frac{1}{16\Gamma t}\right)^d \prod_{\alpha=1}^{d}{\cal F}_{1,\alpha}(v,\sigma_{\alpha},\nu_{l}=0)
 &\mbox{si $r_{\alpha}\ll L$}\\
\kappa_{l}\left(1+\frac{1}{16\Gamma t}\right)^d \prod_{\alpha=1}^{d}{\cal F}_{2,\alpha}(v,\sigma_{\alpha},\nu_{l}=0)
 &\mbox{si $r_{\alpha}\sim L\gg 1$}  ,
\end{array}
\right.
\end{eqnarray}
where the quantities ${\cal F}_{1,\alpha}$ and  ${\cal F}_{2,\alpha}$ are obtained, respectively, from (\ref{eq.5.14.1}) and (\ref{eq.5.14.2}) on substitution of $\sigma$ by $\sigma_{\alpha}\equiv r_{\alpha}/L$.
\section{The Mapping onto a RSOS growth model with ``three states''}
In this section, we introduce a growth model of 
RSOS type ({\it Random Solid on Solid }) with ``three states'', by exploiting a
mapping of the one dimensional model studied in the previous sections.

Let us briefly recall that the RSOS growth models are e.g. useful to describe the spatial fluctuations of the (one-dimensional, of length $L$) interface location in the magnetization profile between coexistent phases in two-dimensional models of ferromagnets, such as in the zero field planar Ising model \cite{Schutz0}. In such models, at zero temperature, every path minimizing the energy of the systen is a sequence of $L$ binary numbers $n_{j}=0,1$ with $j=1,\dots,L$. The stochastic variable $n_{j}$ has the value $n_{j}=0$ if the $j^{th}$ segment of the interface steps upwards (in an angle of $\pi/4$). The value $n_{j}=1$ corresponds to the case where the segments steps downwards with an angle of $\pi/4$. The quantities $n_{j}=0,1$ can be interpreted as occupation numbers relating the interface height $h_{j}$ according to $h_{j}-h_{j-1}=1-2n_{j}$\cite{Schutz0}. In this case the displacement $\delta h_{r}(t)$, at time $t$, of the segment of the interface from the sites $j_{1}$ to $j_{2}>j_{1}$, with $r\equiv j_{2}-j_{1}>0$ is given by $\delta h_{r}(t)=\sum_{k=1}^{r}(1-2n_{k}(t))$ and thus $|\delta h_{r+1}-\delta h_{r}|=1$.

Here we consider an extension of the above model. We consider that the configurations minimizing the energy (at zero temperature) are of the form $\left\{ x_{A}n_{1}^{A}+x_{B}n_{1}^{B},x_{A}n_{2}^{A}+x_{B}n_{2}^{B}, \dots, x_{A}n_{L}^{A}+x_{B}n_{L}^{B} \right\}$, where the discrete stochastic variable  $x_{A}n_{j}^{A}+x_{B}n_{j}^{B}$ can take {\it three} values. The case $x_{A}n_{j}^{A}+x_{B}n_{j}^{B}=0$ again corresponds to the situation where the $j^{th}$ segment steps upwards with an angle of $\pi/4$. The case $x_{A}n_{j}^{A}+x_{B}n_{j}^{B}=x_{A}$ (respectively, $x_{A}n_{j}^{A}+x_{B}n_{j}^{B}=x_{B}$) describe the situation where the $j^{th}$ segment forms an angle $Arctan(1-x_{A})$ (respectively,  $Arctan(1-x_{B})$) with the horizontal. When $x_{B}=0$ (respectively $x_{A}=0$) and $x_{A}=2$ (respectively $x_{B}=2$), we recover the above-mentioned two-state RSOS growth model. In this sense the growth model which we study hereafter is a ``three-states'' extension of the usual RSOS model \cite{Schutz0,Episov,Grynberg,Grynberg1}. In addition the mapping with the diffusive model is clear: the presence of particle of species $A$ (resp. $B$) at site $j$ translates in the language of the growth model with the fact that the related segment of the interface forms  an angle  $Arctan(1-x_{A})$ (respectively,  $Arctan(1-x_{B})$) with the horizontal. In this picture, the  jumping of the diffusive particles correponds to the fluctuation of the orientation  of the related segments of the interface.

We consider a translationally invariant system; the displacement of the (one-dimensional) interface, at time $t$, from the sites $k_{1}$ to $k_{2}>k_{1}$, with $r=k_{2}-k_{1}>0$ is given by  $h_{k_{2}}(t)- h_{k_{1}}(t)\equiv \delta h_{r}(t)$, where
\begin{eqnarray}
\label{eq.5.14}
\delta h_{r}(t)\equiv \sum_{m=1}^{r}\left(1-x_{A}n_{m}^{A}-x_{B}n_{m}^{B}\right)
\end{eqnarray}
Therefore, in the model considered here, we have $max|\delta h_{r+1}(t)-\delta h_{r}(t) |= max(|x_{A}-1|,|x_{B}-1|, |x_{A}+x_{B}-1|)$, instead of the usual constraint  $|\delta h_{r+1}(t)-\delta h_{r}(t)|=1$ of the conventional RSOS models. The mean-displacement of the interface reads $\langle \delta h_{r}(t)\rangle=r(1-x_{A}\rho_{A}-x_{B}\rho_{B})$ thus , if one wants to impose a zero mean-displacement of the interface , we have to require  that $x_{A}\rho_{A}+x_{B}\rho_{B}=1$.

In this section we are interested in the computation of the fluctuations of $\delta h_{j}(t)$: 
\begin{eqnarray}
\label{eq.5.15}
\langle(\delta h_{r}(t))^{2} \rangle\equiv w^{2}(r,t)=
\left((x_{A}+x_{B})^{2}-1\right)r^{2}+\sum_{r'=1}^{r}\sum_{l=1}^{r-r'+1}
\left[x_{A}^{2}\langle n_{l}^{A}n_{l+r'}^{A}\rangle (t)
+x_{B}^{2}\langle n_{l}^{B}n_{l+r'}^{B}\rangle (t)
+2x_{A}x_{B}\langle n_{l}^{A}n_{l+r'}^{B}\rangle (t)
\right],
\end{eqnarray}
where $w(r,t)$ is the  physical width  of the interface .

Using the fact that $ \frac{\partial}{\partial t} \sum_{l=1}^{r}\langle n_{m}^{i}n_{m+l}^{j}\rangle(t)=2\Gamma\left(\langle n_{m}^{i}n_{m+r+1}^{j}\rangle(t)
-\langle n_{m}^{i}n_{m+r}^{j}\rangle(t)
\right)$ and $\sum_{s=1}^{r}\sum_{l=1}^{r-s}\frac{\partial}{\partial t}
\langle n_{m}^{i}n_{m+l}^{j}\rangle(t)=
2\Gamma\left(\langle n_{m}^{i}n_{m+r+1}^{j}\rangle(t)
-\langle n_{m}^{i}n_{m+1}^{j}\rangle(t)
\right)$, we obtain the following equation of motion for the width:
\begin{eqnarray}
\label{eq.5.16}
\frac{\partial w^{2}(r,t)}{\partial t}=4\Gamma\left[x_{A}^2\left\{{\cal C}_{r}^{AA}(t) -{\cal C}_{1}^{AA}(t)\right\}
+ x_{B}^2 \left\{{\cal C}_{r}^{BB}(t) -{\cal C}_{1}^{BB}(t)\right\}
+ 2x_{A}x_{B} \left\{{\cal C}_{r}^{AB}(t) -{\cal C}_{1}^{AB}(t)\right\}
\right],\;r> 1,
\end{eqnarray}
with $w^{2}(1,t)=w^{2}(1,t=0)=(x_{A}+x_{B})^{2}+1-2(\rho_{A}x_{A}+\rho_{B}x_{B})$.

For $r>1$, we thus have:
\begin{eqnarray}
\label{eq.5.17}
w^{2}(r,t)&=&w^{2}(r,0)\nonumber\\ &+& 4\Gamma\int_{0}^{t} dt' \left[x_{A}^2\left\{{\cal C}_{r}^{AA}(t') -{\cal C}_{1}^{AA}(t')\right\}
+ x_{B}^2 \left\{{\cal C}_{r}^{BB}(t') -{\cal C}_{1}^{BB}(t')\right\}
+ 2x_{A}x_{B} \left\{{\cal C}_{r}^{AB}(t') -{\cal C}_{1}^{AB}(t')\right\}
\right],,
\end{eqnarray}
where 
\begin{eqnarray}
\label{eq.5.18}
w^{2}(r,0)&=&[(x_{A}+x_{B})^{2}+1-2(\rho_{A} x_{A} +\rho_{B} x_{B})]r^{2}\nonumber\\&+&2\sum_{r'=1}^{r}(r-r')\left[x_{A}^{2}\langle n_{m}^{A}n_{m+r'}^{A}\rangle(0) +x_{B}^{2}\langle n_{m}^{B}n_{m+r'}^{B}\rangle(0) +2x_{A}x_{B}\langle n_{m}^{A}n_{m+r'}^{B}\rangle(0) \right]
\end{eqnarray}
With help of the formula \cite{tables}$: 4\Gamma \int_{0}^{t}dt' e^{-4\Gamma t'}I_{n}(4\Gamma t')=4\Gamma t e^{-4\Gamma t}\left(I_{0}(4\Gamma t) + I_{1}(4\Gamma t)\right)
+n\left(e^{-4\Gamma t}I_{0}(4\Gamma t)-1\right) +2e^{-4\Gamma t}\sum_{k=1}^{n-1}(n-k)I_{k}(4\Gamma t)$,  and with  the explicit expression of the correlation functions (\ref{eq.5.10}) and (\ref{eq.5.10.0}), we obtain the following exact expression for $w(r,t)$:
\begin{eqnarray}
\label{eq.5.19}
w^{2}(r,t)-w^{2}(r,0)&=&2e^{-4\Gamma t}\sum_{r'\geq 1}\left(x_{A}^{2}{\cal C}_{r'}^{AA}(0)+x_{B}^{2}{\cal C}_{r'}^{BB}(0)+2x_{A}x_{B}{\cal C}_{r'}^{AB}(0)
 \right)\nonumber\\ &\times& \left[\sum_{k=1}^{r+r'-2}(r+r'-k-1)I_{k}(4\Gamma t) + \sum_{k=1}^{r+r'-1}(r'-k-r)I_{k}(4\Gamma t)-2\sum_{k=1}^{r'-2}(r'-k)I_{k}(4\Gamma t) -I_{r'-1}(4\Gamma t)\right]
\end{eqnarray}

We will now specifically focus on two kinds of initial states:

(i) We assume first that the system is initially characterized by an 
alternating periodic array of particles of type 
$A$ and $B$. We consider thus an initial state $|P(0)\rangle=\frac{1}{2}\left(|ABAB\dots\rangle +|BABA\dots\rangle \right)$, with $x_{A}+x_{B}=2$ (in this case $\langle\delta h_{r}(t)\rangle=0 $).
This initial {\it flat interface} leads to $\rho_{A}
=\rho_{B}=1/2$, and for the connected initial correlation functions, we have: ${\cal C}_{r}^{AA}(0)=\frac{(-1)^r}{4}=
{\cal C}_{r}^{BB}(0)=-{\cal C}_{r}^{AB}(0)$.

It follows from (\ref{eq.5.18}) that the initial fluctuations read in this case: $w^{2}(r,0)=r\left(4r+\frac{3}{4}(x_{A}-x_{B})^{2}-1\right)$, for $r>1$ and $r$ even. Therefore, we also have $w^{2}(1,t)=3$.

For this initial configuration, the expression (\ref{eq.5.10}) simplifies and we have ${\cal C}_{1}^{AA}(t)={\cal C}_{1}^{BB}(t)=-{\cal C}_{1}^{AB}(t)=-\frac{e^{-4\Gamma t}}{4}I_{0}(4\Gamma t)$. The general expression (\ref{eq.5.19}) of the fluctations reads: $w^{2}(r,t)-w^{2}(r,0)=(x_{A}-x_{B})^{2}\Gamma \int_{0}^{t}dt' e^{-4\Gamma t'}\left[I_{0}(4\Gamma t')+\sum_{r'\geq 1}(-1)^{r'}\left(I_{r+r'-1}(4\Gamma t')+I_{r-r'}(4\Gamma t')\right)\right]$. From this expression, using the asymptotic behavior of the Bessel functions, it is possible to obtain the long-time behavior (for $\Gamma t \gg 1$ $r \gg 1$) of the fluctuations
\begin{eqnarray}
\label{eq.5.20}
w(r,t)^{2}-w(r,0)^{2}&=&(x_{A}-x_{B})^{2}
\sqrt{\frac{\Gamma t}{2\pi}}(1+{\cal O}((\Gamma t)^{-1}))
\nonumber\\&+& (x_{A}-x_{B})^{2}\sum_{r'\geq 1}(-1)^{r'}\int_{0}^{t}\frac{dt'}{\sqrt{8\pi\Gamma t'}}\left\{
e^{-(r+r'-1)^{2}/8\pi\Gamma t'} + e^{-(r-r')^{2}/8\pi\Gamma t'}\right\}
\left(1+{\cal O}((r-r')^{-2})\right)
\end{eqnarray}
Thus for the initial condition considered here, when $x_{A}\neq x_{B}$, it follows from (\ref{eq.5.20}) that the fluctuations growth as $w(r,t)\approx (\Gamma t )^{1/4}$.

On the other hand, it is known that for the {\it flat interface} (or ``sawtooth'' initial state), which dynamics is coded in a two-state model
$\delta h_{r}=\sum_{m=0}^{n}(1-2n_{m}^{A}) $, the fluctuations growth as $\sim (\Gamma t)^{1/4}$ \cite{Episov,Grynberg,Schutz0}. In the situation considered here, the fluctuations still grow as  $w(r,t)\sim (\Gamma t)^{1/4}$ and thus the details of the model and its ``three-state'' character only appears through the amplitude $(x_{A}-x_{B})^{2}$. When $x_{A}=x_{B}=1$, the initial configuration corresponds to a straight line and, according to (\ref{eq.5.20}), there are no fluctuations. Let us also note that when, e.g.,  $x_{A}=2$ and $x_{B}=0$, the $B$ particles play the role of the vacancies in the two-state model and the model (\ref{eq.5.14}) is exactly mapped onto the well studied two-state model \cite{Episov,Grynberg,Schutz0}.

(ii)
We now investigate the fluctuations of $\delta h_{r}(t)$
in the presence of initial correlations.

We assume that
particles of type $A$ and $B$ are distributed according to \cite{Grynberg}
\begin{eqnarray}
\label{eq.5.23}
{\cal C}_{r}^{l}(0)=\kappa_{l} r^{-\nu_{l}}, \nu_{l}\geq 0, l\in(AA, BB, AB) 
\end{eqnarray}
which corresponds, via the mapping (\ref{eq.5.14}), in the language of the growth model, to an interface with initial fluctuations given by (\ref{eq.5.18}),
where $\langle n_{m}^{i}n_{m+r'}^{j}\rangle(0)={\cal C}^{ij}_{r'}(0)+\rho_{i}\rho_{j}, \; (i,j)\in(A,B)$

Let us define the following quantity:
\begin{eqnarray}
\label{eq.5.24}
\nu=min(\nu_{AA}, \nu_{BB}, \nu_{AB})
\end{eqnarray}

With help of (\ref{eq.5.16}) and (\ref{eq.5.11}), (\ref{eq.5.12}), we 
can compute th asymptotic expression of the fluctuations for $\Gamma t\gg 1$ and $r\gg 1$, with $v\equiv\frac{L^{2}}{8\Gamma t}$ and $\sigma\equiv r/L={\cal O}(1)$ (we assume that $x_{A}x_{B}\neq 0$), which reads
\begin{eqnarray}
\label{eq.5.25}
w(r,t)^{2}-w(r,0)^{2}\sim \left\{
\begin{array}{l l l}
\frac{2\Gamma t}{\sqrt{\pi}\nu(8\pi\Gamma t)^{\nu/2}}\left({\cal F}_{1}(v,\sigma,\nu)-{\cal F}_{2}(v,\sigma,\nu) \right)  &\mbox{, if $0<\nu_{l}<1$}\\
\sqrt{2\Gamma t}\xi(\nu)(1-e^{-\sigma^{2}v})   
&\mbox{, if $\nu_{l}>1$ }\\
\sqrt{\frac{2\Gamma t}{\pi}}\left(\ln(8\Gamma v t)-e^{-\sigma^{2}v}\ln(8\Gamma v \sigma t)\right)
&\mbox{, if $\nu_{l}=1$ } 
\end{array}
\right.
\end{eqnarray}
where the quantities ${\cal F}_{1}$ and ${\cal F}_{2}$ have been defined in (\ref{eq.5.14.1}),(\ref{eq.5.14.2}) and $\xi(\nu)$ is the usual Riemann zeta function.

We see that in the presence of initial correlations (\ref{eq.5.23}), the fluctuations are dominated by the smallest initial correlation exponent $\nu$. Therefore, if $\nu=\nu_{AA}$ (resp. $\nu=\nu_{BB}$ ), the dominant contribution (\ref{eq.5.25}) to the fluctuations are the same as for a correlated two-state model where $x_{A}=2$ and $x_{B}=0$ (resp. $x_{A}=0$ and $x_{B}=2$) where the $B$ (resp. the $A$) particles play the role of vacancies \cite{Grynberg}. 

From (\ref{eq.5.25}) we see that initial correlations affect the long-time behavior of the fluctuations of the height displacement of the interface: when the correlations are ``strong enough'' (i.e. $0<\nu<1$), thus $w(r,t)\sim (\Gamma t)^{\frac{1}{2}-\frac{\nu}{4}}$. Conversely, for ``weak'' initial correlations ($\nu>1$) we recover the usual fluctuation exponent: 
 $w(r,t)\sim (\Gamma t)^{1/4}$. The intermediate case $\nu=1$, corresponds to the {\it marginal} behavior where $w(r,t)\sim (\Gamma t)^{\frac{1}{4}}\sqrt{\ln \Gamma t}$

\section{Non-instantaneous Two-point Correlation functions}
In  this section, we compute exactly the non-instantaneous  two-point correlation functions for various initial states.
Similar quantities have already been computed, for some specific {\it single-species} models (see e.g. \cite{Grynberg,Bennaim,Schutz3,BM1}).

Let us first consider an uncorrelated initial distribution $|P(0)\rangle =
\left(
 \begin{array}{c}
 1-\rho_{A}(0)-\rho_{B}(0)\\
\rho_{A}(0)\\
\rho_{B}(0)
 \end{array}\right)^{\otimes L^{d}}$, 
such that
\begin{eqnarray}
\label{eq.5.27}
\langle n_{{\bf m}}^{A}(0) n_{{\bf l}}^{A}(0)\rangle&=& \rho_{A}(0)\delta_{{\bf m},{\bf l}} +
\rho_{A}(0)^{2}(1-\delta_{{\bf m},{\bf l}})\;\;;\;\;
\langle n_{{\bf m}}^{B}(0) n_{{\bf l}}^{B}(0)\rangle=\rho_{B}(0)\delta_{{\bf m},{\bf l}} +
\rho_{B}(0)^{2}(1-\delta_{{\bf m},{\bf l}});\nonumber\\
\langle n_{{\bf m}}^{A}(0) n_{{\bf l}}^{B}(0)\rangle &=&\langle n_{{\bf m}}^{B}(0) n_{{\bf l}}^{A}(0)\rangle=
 \rho_{A}(0)\rho_{B}(0) (1-\delta_{{\bf m},{\bf l}})
\end{eqnarray}
We then have :
\begin{eqnarray}
\label{eq.5.28}
\langle n_{\bf{m}}^{A}(t) n_{\bf{l}}^{A}(0)\rangle =
\rho_{A}^2(0)
+ (\rho_{A}(0)-\rho_{A}^2(0))\prod_{\alpha=1\dots d}
e^{-2\Gamma t}I_{m_{\alpha}-l_{}\alpha}(2\Gamma t)
\end{eqnarray}
\begin{eqnarray}
\label{eq.5.29}
\langle n_{\bf{m}}^{A}(t) n_{\bf{l}}^{B}(0)\rangle=
\rho_{A}(0)\rho_{B}(0)\left[
1- \prod_{\alpha=1\dots d}
e^{-2\Gamma t}I_{m_\alpha{}-l_{\alpha}}(2\Gamma t)\right]= \langle n_{\bf{m}}^{B}(t) n_{\bf{l}}^{A}(0)\rangle
\end{eqnarray}
\begin{eqnarray}
\label{eq.5.30}
\langle n_{\bf{m}}^{B}(t) n_{\bf{l}}^{B}(0)\rangle =
\rho_{B}^2(0) 
+ (\rho_{B}(0)-\rho_{B}^2(0))\prod_{\alpha=1\dots d}
e^{-2\Gamma t}I_{m_{\alpha}-l_{\alpha}}(2\Gamma t),
\end{eqnarray}
 We  are interested in the asymptotic behavior ($\Gamma t \gg 1$ and  $u=L^2/4\Gamma t <\infty$) of the above functions in the two regimes:
\\
i) $|m_{\alpha}-l_{\alpha}|\equiv r_{\alpha}\sim L \gg 1$, in this case $\sigma_{\alpha}=r_{\alpha}/L={\cal O}(1)$.\\
ii) $|m_{\alpha}-l_{\alpha}|\equiv r_{\alpha}\ll 1$, in this case $\sigma_{\alpha}=r_{\alpha}/L={\cal O}(1/L)$.

It is worth noting that the 
autocorrelation functions are obtained in the second regimes (ii).
We then have
\begin{eqnarray}
\label{eq.5.32}
\langle n_{\bf{m}}^{A}(t) n_{\bf{l}}^{A}(0)\rangle =
\rho_{A}^2(0)  + \frac{(\rho_{A}(0)-\rho_{A}^2(0)) e^{-\sum_{\alpha=1}^{d}\sigma_{\alpha}^2 u} }{(8\pi \Gamma t)^{d/2}} +{\cal O}(1/t^d)
\end{eqnarray}
\begin{eqnarray}
\label{eq.5.33}
\langle n_{\bf{m}}^{A}(t) n_{\bf{l}}^{B}(0)\rangle =
\rho_{A}(0)\rho_{B}(0)\left[1 - \frac{ e^{-\sum_{\alpha=1}^{d}\sigma_{\alpha}^2 u }}{(4\pi \Gamma t)^{d/2}}
 +{\cal O}(1/t^d)\right]=\langle n_{\bf{m}}^{B}(t) n_{\bf{l}}^{A}(0)\rangle
\end{eqnarray}
\begin{eqnarray}
\label{eq.5.34}
\langle n_{\bf{m}}^{B}(t) n_{\bf{l}}^{B}(0)\rangle &=& 
\rho_{B}^2(0)
+\frac{(\rho_{B}(0)-\rho_{B}^2(0)) e^{-\sum_{\alpha=1}^{d}\sigma_{\alpha}^2 u }  }{(4\pi \Gamma t)^{d/2}} +{\cal O}(1/t^d)
\end{eqnarray}
In these regimes, we have a power-law decay of correlation functions $(i,j)\in (A,B) $), namely,
\begin{eqnarray}
\label{eq.5.36}
\langle n_{\bf{m}}^{i}(t) n_{\bf{l}}^{j}(0)\rangle\sim 
(\Gamma t)^{-d/2}   e^{-\sum_{\alpha=1}^{d}\sigma_{\alpha}^2 u}
\end{eqnarray}

Let us now pass to the case where the initial state is 
correlated according to 
\begin{eqnarray}
\label{eq.5.41.0}
\langle n_{\bf{m}}^{i}(0) n_{\bf{l}}^{j}(0)\rangle={\cal K}_{ij}\prod_{\alpha=1,\dots, d}(1-\delta_{r_{\alpha},0})|r_{\alpha}|^{-\Delta_{ij}^{\alpha}},\; 
r_{\alpha}\equiv |m_{\alpha}-l_{\alpha}| ,\;\Delta_{ij}^{\alpha}>0,\; (ij)\in (A,B), \;{\cal K}_{ij}>0, dist(l,m)>0.
\end{eqnarray}
and
\begin{eqnarray}
\label{eq.5.41.1}
\langle n_{\bf{m}}^{i}(0) n_{\bf{m}}^{j}(0)\rangle =\rho_{i}(0)\rho_{j}(0)\delta_{ij}
\end{eqnarray}
Notice that in one dimension the initial state (\ref{eq.5.41.0},\ref{eq.5.41.1}) is translationally invariant and  reads:
\begin{eqnarray}
\label{eq.5.42.0}
\langle n_{m}^{i}(0) n_{l}^{j}(0)\rangle=
\langle n_{ r=|m-l|}^{i}(0) n_{0}^{j}(0)\rangle=
{\cal K}_{ij}(1-\delta_{ r,0})r^{-\Delta_{ij}}+
\rho_{i}(0)\delta_{i,j}\delta_{r,0}; \;\; {\cal K}_{ij}={\cal K}_{ji}; \;\;
 \triangle_{ij}=\triangle_{ji}
\end{eqnarray}
This translational invariance which is broken (see (\ref{eq.5.41.0})) in higher ($d\geq 2$) dimensions leads to  two regimes:

i) We begin with the one-dimensional case ($d=1$), 
here $r=r_{\alpha}\equiv |m-l|$. Because of the initial translational invariant state, we expect that the non-instantaneous correlation functions only depends on $r=m-l$, and we obtain
\begin{eqnarray}
\label{eq.5.43.0}
\langle n_{m}^{A}(t) n_{l}^{A}(0)\rangle=\langle n_{r}^{A}(t) n_{0}^{A}(0)\rangle = 
\rho_{A}(0) e^{2\Gamma t}I_{r}(2\Gamma t)+
{\cal K}_{AA}\sum_{r'\neq 0} |r'|^{-\Delta_{AA}} e^{-2\Gamma t} 
I_{r-r'}(2\Gamma t)
\end{eqnarray}
\begin{eqnarray}
\label{eq.5.44.0}
\langle n_{m}^{A}(t) n_{l}^{B}(0)\rangle =
{\cal K}_{AB}\sum_{r'\neq 0} 
 |r'|^{-\Delta_{AB}} e^{-2\Gamma t} I_{r-r'}(2 \Gamma t)=
\langle n_{m}^{B}(t) n_{l}^{A}(0)\rangle=\langle n_{m}^{B}(t) n_{l}^{A}(0)\rangle
\end{eqnarray}
\begin{eqnarray}
\label{eq.5.45.0}
\langle n_{m}^{B}(t) n_{l}^{B}(0)\rangle=\langle n_{r}^{B}(t) n_{0}^{B}(0)\rangle  =
\rho_{B}(0) e^{-2\Gamma t} I_{r}(2 \Gamma t)+
{\cal K}_{BB}\sum_{r'\neq 0} |r'|^{-\Delta_{BB}} e^{-2\Gamma t} 
I_{r-r'}(2 \Gamma t)
\end{eqnarray}
 Because all the processes in the evolution operator are symmetric (unbiased), there is no drift therefore
 $\langle n_{r}^{A,B}(t) n_{0}^{A,B}(0)\rangle= \langle n_{-r}^{A,B}(t)n_{0}^{A,B}(0)\rangle $.

ii) In higher dimensions ($d\geq 2$),
\begin{eqnarray}
\label{eq.5.42}
\langle n_{\bf{m}}^{A}(t) n_{\bf{l}}^{A}(0)\rangle =
\rho_{A}(0) e^{-2d\Gamma t}\prod_{\alpha=1\dots d}I_{m_{\alpha}-m'_{\alpha}}(2\Gamma t) +
{\cal K}_{AA}\sum_{(m_{1}'\neq l_{1},\dots, m_{d}'\neq l_{d})} \prod_{\alpha=1\dots d} |m'_{\alpha}-l_{\alpha}|^{-\Delta_{AA}^{\alpha} } e^{-2\Gamma t} I_{m_{\alpha}-m'_{\alpha}}(2 \Gamma t)
\end{eqnarray}
\begin{eqnarray}
\label{eq.5.43}
\langle n_{\bf{m}}^{A}(t) n_{\bf{l}}^{B}(0)\rangle =
{\cal K}_{AB}\sum_{(m_{1}'\neq l_{1},\dots, m_{d}'\neq l_{d})} \prod_{\alpha=1\dots d} |m'_{\alpha}-l_{\alpha}|^{-\Delta_{AB}^{\alpha} } e^{-2\Gamma t} I_{m_{i}-m'_{i}}(2 \Gamma t)= \langle n_{\bf{m}}^{B}(t) n_{\bf{l}}^{A}(0)\rangle
\end{eqnarray}
\begin{eqnarray}
\label{eq.5.44}
\langle n_{\bf{m}}^{B}(t) n_{\bf{l}}^{B}(0)\rangle =
\rho_{B}(0) e^{-2d\Gamma t}\prod_{\alpha=1\dots d}I_{m_{\alpha}-m'_{\alpha}}(2\Gamma t) +
{\cal K}_{BB}\sum_{(m_{1}'\neq l_{1},\dots, m_{d}'\neq l_{d})} \prod_{\alpha=1\dots d} |m'_{\alpha}-l_{\alpha}|^{-\Delta_{BB}^{\alpha} } e^{-2\Gamma t} I_{m_{\alpha}-m'_{\alpha}}(2 \Gamma t)
\end{eqnarray}
\begin{eqnarray}
\label{eq.5.45}
\langle n_{\bf{m}}^{B}(t) n_{\bf{l}}^{A}(0)\rangle =
{\cal K}_{BA}\sum_{(m_{1}'\neq l_{1},\dots, m_{d}'\neq l_{d})} \prod_{\alpha=1\dots d} |m'_{\alpha}-l_{\alpha}|^{-\Delta_{BA}^{\alpha} } e^{-2\Gamma t} I_{m_{i}-m'_{i}}(2 \Gamma t)
\end{eqnarray}
We observe that in higher dimensions, because of the initial correlations, 
the non-instantaneous  correlation functions no longer depend on  $|m_{\alpha}-l_{\alpha}|$.

We can express the asymptotic behavior of these non-instantaneous correlation functions in an 
unified way including both $d=1$ and $d\geq 2$ cases.
Assuming that  
$r_{\alpha}=|m_{\alpha}-l_{\alpha}|\sim|m_{\alpha}|\gg 1$, with
$r_{\alpha}=\sigma_{\alpha}L$ and $u =L^2/4\Gamma t <\infty$,  
$\Gamma t, r \gg1$, the asymptotics reads ($(i,j) \in (A,B)$)
\begin{eqnarray}
\label{eq.5.46}
\langle n_{\bf{m}}^{i}(t) n_{\bf{l}}^{j}(0)\rangle =
\left(\frac{\rho_{i}(0)   e^{-\sum_{\alpha=1}^{d}\sigma_{\alpha}^2 u } \delta_{i,j}}{(4\pi \Gamma t)^{d/2}}  
+{\cal K}_{ij} \prod_{\alpha=1\dots d}
 \left[\frac{e^{-\sigma_{\alpha}^{2} u}}{1-\Delta_{ij}^{\alpha}}\sqrt{\frac{u\sigma_{\alpha}^{2}}{\pi}}\frac{1}{4u \Gamma \sigma_{\alpha}^{2}t^{\Delta_{ij}^{\alpha}/2}}\right] + {\cal O}(t^{-2d})\right)\;,\;
 0<\Delta_{ij}^{\alpha}<1
\end{eqnarray}
Moreover 
\begin{eqnarray}
\label{eq.5.47}
\langle n_{\bf{m}}^{i}(t) n_{\bf{l}}^{j}(0)\rangle =
\frac{1}{(4\pi\Gamma t)^{d/2}} \left(\rho_{i}(0)   e^{-\sum_{\alpha=1}^{d}\sigma_{\alpha}^2 u } \delta_{i,j} + {\cal K}_{ij}
 \prod_{\alpha=1\dots d} \zeta(\Delta_{ij}^{\alpha} ) + {\cal O}(t^{-2d})\right)\;,\;
 \Delta_{ij}^{\alpha}>1
\end{eqnarray}
When $\Delta_{ij}^{\alpha}=1$, we have the marginal case with logarithmic
corrections
\begin{eqnarray}
\label{eq.5.48}
\langle n_{\bf{m}}^{i}(t) n_{\bf{l}}^{j}(0)\rangle =
\frac{1}{(4\pi \Gamma t)^{d/2}} \left(\rho_{i}(0)  e^{-\sum_{\alpha=1}^{d}\sigma_{\alpha}^2 u }  \delta_{i,j}  + {\cal K}_{ij}
 \prod_{\alpha=1\dots d} \ln{(4u \sigma_{\alpha} \Gamma t)} + {\cal O}(t^{-2d}) )\right)\;,\;
 \Delta_{ij}^{\alpha}=1
\end{eqnarray}
Again, strong initial correlations lead to an algebraic decay of correlation
functions
($0<\Delta_{ij}^{\alpha}<1 $), 
$\langle n_{\bf{m}}^{i}(t) n_{\bf{l}}^{j}(0)\rangle \sim
 \frac{ 1 }{(4\pi \Gamma t)^{\sum_{\alpha}\Delta_{ij}^{\alpha} /2}}, 
 \;\;0<\Delta_{ij}^{\alpha}<1$,
while for weak initial correlations, we have $\langle n_{\bf{m}}^{i}(t) n_{\bf{l}}^{j}(0)\rangle \sim
 \frac{1}{(4\pi \Gamma t)^{d/2}} \;, \; \Delta_{ij}^{\alpha}>1$.
The marginal case $\Delta_{ij}^{\alpha}=1$ is characterized by
$\langle n_{\bf{m}}^{i}(t) n_{\bf{l}}^{j}(0)\rangle \sim \frac{(\ln{4\Gamma t})^d}{(4\pi \Gamma t)^{d/2}}
 \;, \; \Delta_{ij}^{\alpha}=1 
$
\section{Summary and Conclusion}
 In this work we studied, by analytical methods, the dynamics of a symmetric two-species reaction-diffusion model in arbitrary dimensions. We mapped this model onto an one-dimensional RSOS-type growth model and obtained explicit  results for the latter. In particular, we were able to compute
 the density profile for three various initial conditions (uniform and non-uniform) in arbitrary dimensions.

Furthermore, we evaluated, for a translationally invariant system, the instantaneous two-point correlation functions in arbitrary dimensions.
In one-spatial dimension, we considered the case where initial correlations were present. We observed that when the initial correlations are {\it strong enough}, they affect the asymptotic dynamics. We also noticed a crossover in the dynamics between the case of {\it strong} and {\it weak} initial correlations. 

We mapped the one-dimensional version of the reaction-diffusion model on a ``three-states'' RSOS-type growth model. Using the exact, instantaneous correlation functions, we computed the exact expression of  fluctuations of the interface for the latter model.  We specifically considered the case of a ``flat interface'' where the fluctuations growth as $(\Gamma t)^{1/4}$ , as in the corresponding ``two-state'' growth model: the three-state nature of the model considered only appears in the amplitude. We also considered the case where the initial configuration is translationally-invariant, random and correlated. We saw that the initial correlations are ``strong'', they  affect the long-time behavior of the fluctuations of the displacement interface. Conversely, ``weak'' initial correlations, do not affect the dominant term and the fluctuations still grow as $(\Gamma t)^{1/4}$.
 This is  analog to what happens in  the two-state RSOS systems where initial correlations affect the long-time behaviour of the width \cite{Grynberg}.

Finally we computed  in arbitrary dimensions the  exact non-instantaneous two-point correlation functions  for initially uncorrelated  states
as well as for cases where correlations were present. Here we again observed the effect of {\it strong} initial correlations on the dynamics and a crossover between regimes with {\it strong} and {\it weak} initial correlations takes place.

We conclude this work by addressing an interesting question based on the similarity of the {\it stochastic Hamiltonian }  under consideration with the  integrable Sutherland's quantum spin system  \cite{Alcaraz,Sutherland}. It is known that for one-dimensional SEP-model, the relation with the Heisenberg chain has been fruitful to obtain a {\it (dynamical) matrix formulation} of the probability distribution, which allowed to solve the {\it (dynamical)} density profile for a SEP-model with open boundary conditions (particles were injected and ejected from  both ends of the chain) \cite{Stinchcombe0}. 
The relation of the present model to  Sutherland's one suggests that  a {\it dynamical matrix approach} would be possible (in one-spatial dimension) to treat the injection and ejection of particles at the boundary (open boundary conditions).
\section{Acknowledgments}
The support of Swiss National Fonds is gratefully acknowledged.
\section{Appendix: Solution of the equations of motion of the instantaneous two-point correlation functions}
Solving the equations of motions (\ref{eq.5.8}) for the instantaneous correlation functions in arbitrary dimensions, for the model under consideration in section $VI$, we obtain the following expression:
\begin{eqnarray}
\label{eq.A.1}
&&{\cal C}_{|r|}^{ij}(t)={\cal C}_{0}^{ij}(0) e^{-4d\Gamma t}\prod_{\alpha=1}^{d}
I_{r_{\alpha}}(4\Gamma t)+\sum_{r'\neq 0}{\cal C}_{r'}^{ij}(0) 
e^{-4\Gamma dt}\prod_{\alpha=1}^{d} I_{r_{\alpha}-r'_{\alpha}}(4\Gamma t) \nonumber\\
&+& 4d\Gamma {\cal C}_{0}^{ij}(0) \int_{0}^{t} dt' e^{-4\Gamma d(t-t')}
\prod_{\alpha=1}^{d}I_{r_{\alpha}}(4\Gamma(t-t'))
\nonumber\\
 &-&2\Gamma {\cal C}_{0}^{ij}(0) \int_{0}^{t} dt' e^{-4\Gamma d(t-t')} \sum_{\alpha=1}^{d}\left(\prod_{\alpha\neq\alpha'=1\dots d} I_{r_{\alpha'}}(4\Gamma(t-t'))( I_{r_{\alpha}+1}(4\Gamma(t-t'))+  I_{r_{\alpha}-1}(4\Gamma(t-t'))  )\right) \nonumber\\
 &+&2\Gamma  \int_{0}^{t} dt' e^{-4\Gamma d(t-t')}  \sum_{\alpha=1}^{d}{\cal C}_{e^{\alpha}}^{ij}(t')\left(\prod_{\alpha\neq\alpha'=1\dots d} I_{r_{\alpha'}}(4\Gamma(t-t'))( I_{r_{\alpha}+1}(4\Gamma(t-t')) + I_{r_{\alpha}-1}(4\Gamma(t-t')))\right) \nonumber\\
&-& 4d{\Gamma} \int_{0}^{t} dt' e^{-4\Gamma d(t-t')} {\cal C}_{e^{\alpha}}^{ij}(t')\prod_{\alpha=1}^{d} I_{r_{\alpha}}(4\Gamma(t-t'))
\end{eqnarray}
Using the properties of the derivatives of Bessel functions and then integrating by parts, we obtain the more compact form (\ref{eq.5.9})

\begin{thebibliography}{99}
%
%
\bibitem{Privman}
``Nonequilibrium Statistical Mechanics in One Dimension'', edited by V. Privman (Cambridge University Press, Cambridge, 1997); J. Marro, R. Dickman, ``Phase Transitions in Lattice Systems'' (Cambridge University Press, Cambridge, 1998); B. Chopard, M. Droz, `` Cellular Automata Modelling of Physical Systems'' (Cambridge University Press, Cambridge, 1998); 
D.C. Mattis and M.L. Glasser, Rev. Mod. Phys. {\bf 70}, 979 (1998)
%
\bibitem{Schutz0}
G.M.\,Sch\"{u}tz, ``Exactly solvable models for many-body systems far from equilibrium'' in  Phase Transitions and Critical Phenomena, \,C. Domb and J. Lebowitz (eds.), (Academic Press, London, 2000).
%
\bibitem{Schadschneider}
A. Schadschneider and M. Schreckenberg, J. Phys. A {\bf 26}, L679 (1993); S. Yukawa, M. Kikuchi and S. Tadaki, J. Phys. Soc. Jap. {\bf 63}, 3609 (1994); M. Schreckenberg, A. Schadschneider, K. Nagel and N. Ito, Phys. Rev. E {\bf 51}, 2939 (1995); T. Nagatani, J. Phys. A {\bf 28}, 7079 (1995)
%
\bibitem{Macdonald}
J.T. MacDonald, J.H. Gibbs, A.C. Pipkin, Biopolymer {\bf 6}, 1(1968); G.M. Sch\"{u}tz, Int. J. Mod. Phys. B {\bf 11}, 197 (1997)
%
\bibitem{Widom}
B. Widom, J.L. Viovy and A.D. D\'efontaines, J.Phys.I (France) {\bf 1}, 1759 (1991); J.M.J van Leeuwen and A. Kooiman, Physica A {\bf 184}, 79 (1992); G.T. Barkema, J.F. Marko and B. Widom, Phys. Rev E {\bf 49}, 5303 (1994); G.T. Barkema, C. Caron  and J.F. Marko, Biopolymers {\bf 38}, 665 (1996); G.T. Barkema, G.M. Sch\"{u}tz, Europhys. Lett. {\bf 35}, 139 (1996); M. Pr\"{a}hofer and H. Spohn, physica A {\bf 233}, 191 (1996)
%
\bibitem{Krug}
J. Krug and H.Spohn, in {\it Solids far from Equilibrium}, ed. C. Godr\`eche, (Cambridge University Press, Cambridge, 1991); T. Halpin-Healey and Y.-C. Zhang, Phys. Rep. {\bf 254}, 215 (1995); D.B. Abraham, T.J. Newman and G.M. Sch\"{u}tz, cond-mat/9404064 (1994); 
%
\bibitem{Episov}
S.E. Episov, T.J. Newman, J. Stat. Phys. \,{\bf 70}, 691 \,\,(1993) 
%
\bibitem{Stinchcombe0}
R.B. Stinchcombe and G.M. Sch\"{u}tz, Phys. Rev. Lett. {\bf 75}, 140 (1995); R.B. Stinchcombe and G.M. Sch\"{u}tz, Europhys. Lett. {\bf 29}, 663 (1995)
%
\bibitem{Krug1}
J. Krug, Phys. Rev. Lett. {\bf 67}, 1882 (1991); M.R. Evans, D.P. Foster, C. Godr\`eche, D. Mukamel,  Phys. Rev. Lett. {\bf 74}, 208 (1995); B. Derrida, E. Domany, D. Mukamel, J. Stat. Phys. {\bf 69}, 667 (1993); B. Derrida, M.R. Evans, J. Physique I {\bf 3}, 311 (1993); 
B. Derrida, M.R. Evans, V. Hakim, V. Pasquier, J. Phys. A: Math. Gen. 
{\bf 26}, 1493 (1993); G.M. Sch\"{u}tz and E. Domany J. Stat. Phys. {\bf 72}, 277 (1993);
B. Derrida Phys. Rep. {\bf 301}, 65 (1998)
%
\bibitem{Krug2}
J. Krug, Adv. Phys. {\bf 46}, 139 (1997)
%
\bibitem{Schutz1}
G.M. Sch\"{u}tz Euro. Phys. J. B {\bf 5}, 277 (1998)
%
\bibitem{Derrida}
B. Derrida, S.A. Janowsky, J.L. Lebowitz, E.R. Speer, J. Stat. Phys. {\bf 73}, 813  (1993); M.R. Evans, D.P. Foster, C. Godr\`eche, D. Mukamel , J. Stat. Phys. {\bf 80}, 69  (1995); P.F. Arndt, T. Heinzel, V. Rittenberg,  J.Phys.A {\bf 31} (1998) 833; P.F. Arndt, T. Heinzel, V. Rittenberg, J.Stat.Phys. {\bf 90} (1998) 783; P.F. Arndt, T. Heinzel, V. Rittenberg,  J. Phys. A {\bf 31}, L45  (1998); V. Karimipour , Phys. Rev. E. {\bf 59}, 205  (1999)
%
\bibitem{Arndt}
P. Arndt, Phys. Rev. Lett. \,{\bf 84}, 814 \,\,(2000)
%
\bibitem{Rajewsky}
N. Rajewsky, T. Sasamoto, E.R. Speer, Physica A \,{\bf 279}, 123 \,\,(2000); see also: T. Sasamoto and D. Zagier, J. Phys. A  \,{\bf 34}, 5033 \,\,(2001)
%
\bibitem{Privman1}
V. Privman, Phys. Rev. A \,{\bf 46}, R6140 \,\,(1992)
%
\bibitem{Belitsky}
M. Bramson and J.L. Lebowitz, Phys. Rev. Lett. \,{\bf 61}, 2397 \,\,(1988); M. Bramson and J.L. Lebowitz, J. Stat. Phys. \,{\bf 62}, 297 \,\,(1991); V. Belitsky, J. Stat. Phys. {\bf 73}, 671 (1993)
%
%
\bibitem{BM}
M. Mobilia and P.-A. Bares,  Phys. Rev. E {\bf 63}, 036121 \,\,(2001) 
%
\bibitem{Grynberg}
 M.D. Grynberg, T.J. Newman, and R.B. Stinchcombe, Phys. Rev. E \,{\bf 50}, 957 \,\,(1994)
%
%
\bibitem{Grynberg1}
M.D. Grynberg, J. Stat. Phys. \,,{\bf 103}, 395 \,\, (2001) 
%
%
\bibitem{Schutz2}
G.M. Sch\"{u}tz, J. Stat. Phys.  {\bf 79}, 243 (1995)
%
%
\bibitem{Fujii}
Y. Fujii and M. Wadati,  J. Phys. Soc. of Japan  {\bf 66}, 3770 (1997)
%
\bibitem{Alcaraz}
F. Alcaraz, M. Droz, M. Henkel, V. Rittenberg, Annals of Physics \,{\bf 230}, 250 \,\,(1994)
%
%
\bibitem{Sutherland}
B. Sutherland  Phys. Rev. B \,
{\bf 12}, 3795 \,\,(1975)
%
%
\bibitem{tables}

M. Abramowitz and I. Stegun, ``Handbook of Mathematical Functions'', Dover, NY (1965)
%
%
\bibitem{Bennaim}
E. Ben-Naim, L. Frachebourg, L Krapivski  Phys. Rev. E \,
{\bf 53}, 3078 \,\,(1996)
%
\bibitem{Schutz3}
G.M. Sch\"{u}tz, Phys. Rev. E\, {\bf 53}, 1475 (1996)
%
\bibitem{BM1}
P.-A. Bares and M. Mobilia,  Phys. Rev. E \,{\bf 59}, 1996 \,\,(1999);
M. Mobilia and P.-A. Bares, Phys. Rev. E \,{\bf 63}, 056112 \,\,(2001)
%
\end{thebibliography}
\end{document}